\documentstyle[11pt, aaspp4]{article}
\def\half{\textstyle {1 \over 2} \displaystyle}

\def\5h{\textstyle {5 \over 2} \displaystyle}
\def\qrt{\textstyle {1 \over 4} \displaystyle}
\def\ltsima{$\;\buildrel < \over \sim \;$}
\def\simlt{\lower.5ex \hbox{\ltsima}}
\def\gtsima{$\;\buildrel > \over \sim \;$}
\def\simgt{\lower.5ex \hbox{\gtsima}}
\begin{document}
\title {Response of circumnuclear water masers to 
luminosity changes in an active galactic nucleus}

\author{David A. Neufeld}
  
\affil{Department of Physics \& Astronomy,  The Johns Hopkins University,
3400 North Charles Street,  Baltimore, MD 21218}

\begin{abstract}

Circumnuclear water masers can respond in two ways
to changes in the luminosity of an active galactic
nucleus.  First, an increase in the X-ray luminosity 
can lead to an increase in the maser emissivity; 
and second, an increase
in the intrinsic bolometric luminosity may result in
a temporary {\it decrease} in the difference between the gas 
and dust temperature and
a consequent {\it decrease} in the maser output.
Whilst the latter effect can occur over a period 
shorter than the thermal timescale, the former effect 
cannot.  Quantitative 
estimates of the response of the water maser emissivity to 
changes in either the X-ray or bolometric luminosity are 
presented, together with estimates  of the relevant
timescales.  Either
mechanism could account for recent observations by 
Gallimore et al.\ which suggest that the water
maser variability in two widely separated regions 
of the circumnuclear gas in NGC 1068 have been
coordinated by a signal from the active nucleus.
For either mechanism, a minimum H$_2$ density 
$\sim 10^8\,\rm cm^{-3}$ is needed to explain the
observed variability timescale.

\end{abstract}


\section{Introduction}

Luminous maser emissions in the $6_{16}-5_{23}$ 
22 GHz line of water have now been detected from more than 
20 active galaxies, a list of which is given in Table 2 of the
recent review by Moran, Greenhill \& Herrnstein (2000). 
In almost every case where the size of the emission 
region has been determined interferometrically, the water 
maser emission has  been found to originate primarily 
within a parsec-  or subparsec-sized region that is 
coincident with the active nucleus of the galaxy, suggesting 
that active galactic nuclei are the energy sources that 
ultimately power the observed water maser emission.
In several sources, the geometry is strongly suggestive
of a circumnuclear accretion disk  (e.g. Miyoshi et al.\
1995).

Theoretical models for the collisional excitation of water masers
(e.g. de~Jong 1973, Neufeld \& Melnick 1991) place requirements
on the physical conditions in the emitting region.  Luminous
water maser emission is favored by molecular hydrogen densities,
$n({\rm H}_2)$, in the range $10^7 - 10^{11} \rm cm^{-3}$,
water abundances, $x({\rm H_2O}) = 
n({\rm H_2O})/n({\rm H_2}) \simgt 10^{-4}$, and gas 
temperatures $\simgt 400$~K.  A minimum 
temperature requirement is imposed both by the 
excitation requirement for the
maser transition and by the fact that water abundances are
significantly enhanced at temperatures $\simgt 400$~K 
by the reaction $\rm O + H_2 \rightarrow OH + H$ followed
by $\rm OH + H_2 \rightarrow H_2O + H$ 
(e.g. Elitzur \& de~Jong 1978; Neufeld, Lepp \& Melnick 1995).
An additional requirement is that the gas temperature must significantly
exceed the temperature of the radiation field at mid- 
to far-infrared wavelengths: thus some energy source that
directly heats the circumnuclear gas in active galaxies is
needed to explain the observed water maser emission.  X-rays 
from the active nucleus (Neufeld, Maloney \& Conger 1994,
hereafter NMC94), 
shock waves within a circumnuclear accretion disk 
(Maoz \& McKee 1998), and viscous heating (Desch, Wallin
\& Watson 1998, hereafter DWW98) have all 
been proposed as the energy source responsible for heating 
the masing gas.

Detailed models for water masers in X-ray heated gas have
been presented by NMC94, who considered the water maser
emission as a function of depth into an X-irradiated
slab.  Those models
showed that the physical and chemical conditions within
X-ray heated gas are conducive to the excitation of water 
masers, and that -- given reasonable assumptions about the 
nature of the circumnuclear emission region -- X-ray heating
could account for the typical luminosities of water masers 
observed in AGN.  Collison and Watson (1995, hereafter CW95) 
subsequently made the important point that for certain parameters the 
maser emission can be significantly enhanced by the absorption 
of non-masing far-infrared water lines by dust, a process not
included by NMC94.

Since AGN are known to exhibit significant and rapid variability 
(e.g.\ Ulrich, Maraschi, \& Urry 1997) at optical, ultraviolet, and
particularly X-ray wavelengths (e.g.\ Mushotzki, Done \& Pounds 1993),
maser models that invoke X-ray heating naturally raise the
question of how the water maser emission would respond to
changes in the incident X-ray and bolometric fluxes.  This question, to be
addressed this Letter, is also strongly motivated by observations of
cooordinated maser variability in NGC 1068 that have 
been reported recently by Gallimore et al. (2000).  
In \S 2 below, I present the results of a general parameter 
study for the equilibrium maser emission from a warm dusty 
medium.  In \S 3, I consider how the maser emission 
responds to changes in the illuminating X-ray flux or in the
dust temperature within the masing medium.  In \S 4, the
results are discussed with specific reference to the
water masers in NGC 1068.

\section{Water maser emission from a warm dusty medium}

The collisional excitation of water masers within 
a warm dusty medium has been modeled by Wallin \& Watson
(1997) in the limit where the medium is 
optically thick to far-infrared continuum radiation.
As previously pointed out by CW95,
this limit often applies to circumnuclear gas in AGN,
given the continuum dust opacities typically present
(Draine \& Laor 1993).  The effect of continuum opacity
is to place a lower limit on the escape probability for
non-masing far-infrared transitions of water, allowing 
maser emission to occur even at great depths within a
circumnuclear disk or gas cloud and enhancing the
cooling rate due to far-infrared water emissions 
(DWW98). 

In the limit where the medium is optically thick to 
far-infrared continuum radiation, the water level populations
are determined solely by the local conditions and are
dependent on the gas temperature, $T_{\rm gas}$,
the dust temperature, $T_{\rm dust}$, the H$_2$ density, 
$n({\rm H_2})$, and the water abundance $x({\rm H_2O}) = 
n({\rm H_2O})/n({\rm H_2})$.  Since water abundances
$\sim 10^{-4}$ are predicted for regions of luminous water
maser emission, I have constructed a grid of models 
for fixed $x({\rm H_2O}) \sim 10^{-4}$ that span a 
broad region in the three dimensional parameter space
defined by $T_{\rm gas}$, $T_{\rm dust}$, and $n({\rm H_2})$.

The water level populations are given by the usual equations
of statistical equilibrium (e.g. CW95), with the rates
for all radiative processes diminished by an escape probability
$\beta_{ij}^c = [2 \alpha_{ij}/ (1  +  2 \alpha_{ij})]
(\ln [e + \pi^{-1/2} \alpha_{ij}^{-1}])^{1/2} $, 
where $\alpha_{ij}$ is the ratio of dust continuum opacity
to line center opacity for the transition from $i$ to $j$
(Hollenbach \& McKee 1979, CW95).  For consistency with
the previous work of CW95, we assume a continuum
absorption coefficient of $10^{-23} n_{\rm H} \rm \, cm^{2}$
for wavelengths $\lambda \le \rm 50 \,\mu m$ and 
 $10^{-23} n_{\rm H} ({\rm 50\,\mu m}/\lambda)^2 \rm \, cm^{2}$
for $\lambda \ge \rm 50 \,\mu m$, where $n_{\rm H}$ is
the density of H nuclei. 
For comparison, the probability of escape 
from a semi-infinite plane parallel slab is
$\beta_{ij} = (4 \pi^{1/2} \tau_{ij} 
[\ln \tau_{ij}]^{1/2})^{-1}$ in the limit where
the line center optical depth $\tau_{ij} \gg 1$
(Hollenbach \& McKee 1979).
Comparing these expressions, we find that dust
absorption dominates the effective escape of 
far-infrared water emissions for dust optical
depths $\ge (8 \pi^{1/2})^{-1} = 0.07$,  corresponding
to column densities of H nuclei, 
$N_{\rm H} \simgt 10^{22} \rm cm^{-2}$. 
Following Neufeld \& Kaufman (1993, hereafter NK93), I solved the 
equations of statistical equilibrium for 179 rotational 
states of ortho-water, adopting the collisional rate
coefficients of Green, Maluendes \& McLean (1993)
for transitions among the
lowest 45 states and a simple extrapolation for transitions 
involving states of higher energy.  The assumed line width,
defined as velocity shift needed to reduce the 
opacity from its peak value by a factor e, was
$1\,\rm km\,s^{-1}$.

Once the statistical equilibrium equations
have been solved, the cooling rate due to
water emissions and the rate of maser emission 
under conditions of saturation may be obtained
straightforwardly (e.g. Neufeld \& Melnick
1991; NK93).  
The results I obtained for the maser 
emissivity are in good agreement with those presented
previously by Wallin \& Watson (1997).  
If the gas is in thermal equilbrium, then the required
heating rate is equal to the cooling rate, which
-- for the conditions of relevance here --
is dominated by water cooling and
gas-grain collisional cooling (Neufeld et al.~1995).
Using the gas-grain
cooling rate adopted by DWW98 and the water cooling rate
obtained above, I computed the equilibrium heating rate, $H$,
and used it in place of $T_{\rm gas}$ as one of the
three dependent variables.  

Figure 1a shows a contour 
plot of the gas temperature, 
$T_{\rm gas}$, and the maser efficiency, $\epsilon^0_{\rm sat}
= h\nu \Phi^0_{\rm sat}/( n_{\rm H} H)$, as function of $T_{\rm dust}$
and $H/n_{\rm H}$.  Here $\Phi^0_{\rm sat}$
is the rate at which maser photons are emitted per unit volume,
the superscript 0 indicating that the results apply to
the case where the dust optical depth is large (Watson \& Wallin 1997)
and the subscript sat indicating that the maser emission is 
assumed to be saturated.  The heating rate, $H$, is defined 
following the notation of Maloney, Hollenbach
\& Tielens (1996, hereafter MHT96) as the heating rate
per hydrogen nucleus (with units erg s$^{-1}$).
The maser efficiency 
$\epsilon^0_{\rm sat}$ is thus the fraction of the input 
power that emerges as 22 GHz maser radiation. 
The results shown in Figure 1a
apply to a fixed molecular hydrogen
density $n({\rm H}_2) = \half n_{\rm H} = 10^9 \,\rm cm^{-3}$. 

As required by the second law of thermodynamics, Figure 1 
shows that $T_{\rm gas} \rightarrow T_{\rm dust}$ in the
limit $H \rightarrow 0$.  There is a minimum value  of the
heating rate (defined by the curve $\epsilon^0_{\rm sat} = 0$)
below which $T_{\rm gas} - T_{\rm dust}$ is
insufficient to support a population inversion .
The vertical line at $H/n_H \sim 1.2 \times 
10^{-28} \rm erg \, cm^3 \,
s^{-1}$ represents the {\it maximum} X-ray heating rate that can be sustained
without the gas being dissociated (NMC94).  Thus the region for
luminous maser action within an X-ray heated
medium lies to the right of the curve $\epsilon^0_{\rm sat} = 0$,
to the left of the line $H/n_H = 1.2 \times 
10^{-28} \rm erg \, cm^3 \,
s^{-1}$, and above the curve $T_{\rm gas} = 400$~K (the latter requirement
being needed for consistency with the large water abundance
of 10$^{-4}$ that has been assumed).  Figure 1b presents 
analogous results to those in Figure 1a, but now $\epsilon^0_{\rm sat}$
and $T_{\rm gas}$ are shown for a
fixed dust temperature (500~K)
as a function of $n({\rm H}_2)$ and $H/{\rm n}_H$.

\section{Response of masers to a change in their environment}

Results shown in Figure 1 apply to conditions of
thermal equilibrium. The effects of changing the heating
rate and dust temperature can now be considered.  The
timescale for the gas to re-establish thermal equilibrium
after a change in the heating rate or dust temperature
is given by $\tau_{\rm therm} = \5h kT / 2H 
=$ $0.027 \, (T/1000\,{\rm K}) \,
(n({\rm H}_2) / 10^9\,{\rm cm}^{-3})^{-1} \,
([H/n_{\rm H}] / 
10^{-28} \rm erg\,cm^3\,s^{-1})^{-1} \, yr.$   Let us 
first consider the effects of changing the heating
rate. On timescales greater than $\tau_{\rm therm}$,  
an increase in the heating rate increases the difference
between the gas and dust temperatures and increases the
rate of maser emission.  The response is then conveniently 
characterized by the logarithmic derivative 
$\partial \ln \Phi^0_{\rm sat} / \partial \ln H = 
1 + (\partial \ln \epsilon^0_{\rm sat}
/ \partial \ln H)$, which is plotted in Figure 2 (dashed curve)
for $n({\rm H}_2) = 10^9 \,\rm cm^{-3}$ 
and $T_{\rm dust} = 500$~K.  Except for small values
of $H$ close to the threshold for maser action, the
derivative $\partial  \ln \Phi^0_{\rm sat}
/ \partial \ln H$ lies within
a factor of 2 of unity, implying that changes in the
heating rate are accompanied by comparable fractional
changes in the maser emission rate.  

The results plotted in
Figure 2 are premised on assumptions of constant
density and water abundance.  The density is only expected
to change on a dynamical timescale, which is typically much longer 
than $\tau_{\rm therm}$.   The molecular fraction,
$2\,n({\rm H}_2) /n_{\rm H}$, only changes on the 
X-ray ionization timescale, which is given (MHT96) by $\sim 4\,
(n({\rm H}_2) / 10^9\,{\rm cm}^{-3})^{-1} \,
([H/n_{\rm H}] / 10^{-28} \rm erg\,cm^3\,s^{-1} )^{-1} \, yr$
and is
a factor $\sim 200$ greater than the cooling timescale.
Even if the molecular fraction is constant, however,
the water abundance 
can change in response to changes in the temperature.  This 
effect, the inclusion of which lies beyond the scope 
of the present study, is unlikely to be important except 
when the gas temperature lies close to the threshold for 
rapid water production ($T_{\rm gas} \sim 400$~K)
or when the X-ray ionization rate lies close
to the threshold for molecular dissociation.

If the heating rate varies, the gas temperature and maser
emissivity $\Phi^0_{\rm sat}$ cannot respond on timescales 
any shorter than  $\tau_{\rm therm}$.  However, $\Phi^0_{\rm sat}$ 
{\it can} vary on timescales shorter than $\tau_{\rm therm}$ 
if the dust temperature changes rapidly.  The thermal 
inertia of the dust grains is negligible, so the timescale 
on which the dust temperature can change is controlled 
by the diffusion time for the far-infrared continuum 
radiation\footnote{This is the shortest timescale on
which the {\it local} infrared radiation field can vary;
note that the infrared flux {\it as viewed from afar} may have
a characteristic variability timescale that is considerably
longer than that of the local infrared radiation as
a result of the different light travel times from
the near and far sides of the irradiated disk or torus.}, 
given by $\tau_{\rm rad} \sim \tau_{\rm FIR} \ell /c \sim 0.005  
[N_{H} / 10^{24} \, \rm cm^{-2}]^2  
\,[n({\rm H}_2) / 10^{9} \, \rm cm^{-3}]^{-1} \, yr$, 
where $\ell$ is the distance below the irradiated surface 
of the cloud and $\tau_{FIR} \sim 
10\,(N_{H} / 10^{24} \, \rm cm^{-3})$ (assumed $\ge 1$) is the
corresponding optical depth at far-infrared wavelengths.  

Under circumstances where the dust temperature varies
on timescales shorter than $\tau_{\rm therm}$ --
as it might in response
to variations in the incident bolometric flux from the
AGN, for example -- the gas temperature remains constant
but the maser emissivity varies as a result of changes in
the difference between the gas and dust temperatures,
$T_{\rm diff} = T_{\rm gas} - T_{\rm dust}$.  The response is now
characterized by the derivative $(\partial \ln \Phi^0_{\rm sat} 
/ \partial \ln T_{\rm dust})_{T_{\rm gas}}$, 
the subscript $T_{\rm gas}$ indicating that the gas 
temperature is held constant. The dotted curve in Figure 2 
shows the value of this derivative, again for the case where 
$n({\rm H}_2) = 10^9 \,\rm cm^{-3}$  and $T_{\rm dust} = 500$~K.
Two features of this curve are particularly noteworthy.
First, the value of $(\partial \ln \Phi^0_{\rm sat} 
/ \partial \ln T_{\rm dust})_{T_{\rm gas}}$ is always 
negative because $T_{\rm diff}$ decreases as $T_{\rm dust}$
increases.  Second, $(\partial \ln \Phi^0_{\rm sat} 
/ \partial \ln T_{\rm dust})_{T_{\rm gas}}$ can become
much larger than unity for small values of $H$.  This
behavior occurs because the maser emissivity is primarily
controlled by $T_{\rm diff}$;
a small fractional 
change in $T_{\rm dust}$ causes a large fractional change
in $T_{\rm diff}$ when $T_{\rm diff} \ll T_{\rm dust}$.

\section{Application to NGC 1068}

The results presented in Figures 1 and 2 can now be
discussed with reference to recent observations of
the Seyfert galaxy NGC 1068.  Gallimore et al. (2000) have 
recently reported that two maser features in NGC 1068, 
symmetrically placed in Doppler velocity about the systemic 
velocity of the host galaxy, flared simultaneously with a rise 
time of at most 22 days. Given the emission region structure 
elucidated by earlier VLBA observations (Greenhill \& Gwinn 1997), the
velocities of the maser features involved suggest that
they are symmetrically located at distance 
$d\sim 0.7$ pc on either side 
of the nucleus. Unless the simultaneity of the flares is entirely
happenstance, they were presumably coordinated by a
signal that propagated from the nucleus.

The results presented in Figure 2 suggest
two possible mechanisms 
for how the maser flares might
be coordinated:  (1) variations in the central X-ray
luminosity, $L_X$, might simultaneously change the heating rates
in the masing regions; or (2) variations in the intrinsic bolometric
luminosity, $L_{\rm bol}$, might simultaneously change
the dust temperature\footnote{For the case of 
saturated masers, an third mechanism is possible:  in principle, the
observed flares might be coordinated by a varying 22 GHz continuum
flux from a compact central source.  Since the maser 
radiation from the flaring regions travels tangentially
toward us, the flux we receive from those regions might 
increase if the radial flux of seed radiation from the
central source diminished.  This explanation, however, 
would require that the maser emission is controlled by
seed radiation from a compact continuum source; in NGC 1068
the observed distribution of maser emission close to the
systemic velocity argues against that being the case.}. 
Unfortunately, the average X-ray and bolometric 
luminosities are rather uncertain for NGC 1068, since 
the central source is obscured even for hard X-rays;
I adopted the estimates of Pier et al.\ (1994), who
compiled observations obtained over a wide range of
wavelengths and assumed that 1\% of the total intrinsic
luminosity is reflected towards us.  These estimates
of the luminosities imply a 1 -- 100~keV X-ray 
flux $\sim 7 \times 10^5 \rm \, erg \, cm^{-2} \, s^{-1}$
at the distance of 
the flaring masers and a characteristic dust temperature 
$\sim (L_{\rm bol} /4 \pi d^2 \sigma)^{1/4} \sim 500$~K.
Expressions given by MHT96 then imply an X-ray heating rate
$H \sim 5.6 \times 
10^{-20}\,(N_{\rm H} / 10^{24} \,\rm cm^{-2})^{-0.9}
\rm \, erg \, s^{-1}$ at column density $N_{\rm H}$
below the irradiated surface of the maser emission region.
For column densities greater than $N_{\rm H} \sim 10^{24}\, \rm
cm^{-2}$,
Thomson scattering becomes important and
increases the effective column density
appearing in the above equation for $H$ by a factor
$\sim \tau_{T} = 6.65 \times 10^{-25}\, N_{\rm H} \, \rm cm^2$.

In Figure 3, relevant constraints are 
plotted\footnote{The results plotted here apply to a
gas cloud that is illuminated at normal incidence
such that the column densities for X-ray shielding
and for the escape of far-infrared radiation are
identical.  For oblique illumination at angle $\theta$
to the normal, the column density $N_{\rm H}$ 
refers to the shielding column density, the 
timescale $\tau_{rad}$ is reduced
by a factor sec$^2\theta$, and the dust temperature should be
reduced by a factor $(\rm sec \,\theta)^{1/4}$.  An
additional assumption is that dust absorption dominates the
removal of far-infrared line photons, requiring 
$N_{\rm H} \rm \, cos\, \theta \simgt 10^{22}
\, cm^{-2}$.}  
as
a function of $n({\rm H}_2)$ and $N_{\rm H}$.  Molecules
are present above and to the right of the line 
$H/n_{\rm H} = 1.2 \times 10^{-28} \rm erg\,cm^3\,s^{-1}$.
Maser action can occur below and to the left of the
curve $\epsilon^0_{\rm sat} = 0$.  The observed
variability can be driven by changes in the X-ray
heating rate for parameters to the left of the
line $\tau_{\rm therm} = 0.05$~yr; or alternatively
by changes in the dust temperature for the region
to the left of the line $\tau_{\rm rad} = 0.05$~yr and to the
right of the line $\tau_{\rm therm} = 0.05$~yr.
In either case, the minimum required H$_2$ density 
for a variability timescale of 0.05~yr is $\sim \rm 10^8 
\, cm^{-2}$.
In Figure 4, the derivatives 
$\partial \ln \Phi^0_{\rm sat} / \partial \ln L_X$
and $(\partial \ln \Phi^0_{\rm sat} 
/ \partial \ln L_{\rm bol})_{T_{\rm gas}}$
are shown for several values of the density.
Since the dust temperature increases as $L_{\rm bol}^{1/4}$,
the latter derivative is given by $\qrt (\partial \ln \Phi^0_{\rm sat} 
/ \partial \ln T_{\rm dust})_{T_{\rm gas}}$.

Figures 3 and 4 indicate that the simultaneous flaring of the
two maser features in NGC 1068
could plausibly be coordinated {\it either} through 
variations in the heating rate {\it or} 
through variations in the dust 
temperature.  The former mechanism would be unique to 
X-ray heating whereas the latter might apply in principle
even if shock heating or viscous heating were responsible
for the maser emission.  Observationally, the mechanisms
might be discriminated by searching for correlations 
between the maser emission and X-ray and bolometric
luminosities, although in the case of NGC 1068 the
fact that we observe only reflected radiation from the
nucleus will introduce uncertain delays that will be
difficult to remove.  Maser variability induced by 
a varying X-ray heating rate will be characterized by
a positive correlation with $L_X$ (with a time delay
determined by the different light travel times, of
course), while variability induced by 
a varying dust temperature will be characterized by
a negative correlation with $L_{\rm bol}$.

It is a pleasure to acknowledge helpful discussions
with E.~Agol and J.~Gallimore.
I gratefully acknowledge the support of a National
Young Investigator award from the National Science
Foundation.

\clearpage

\begin{figure}
\plotfiddle{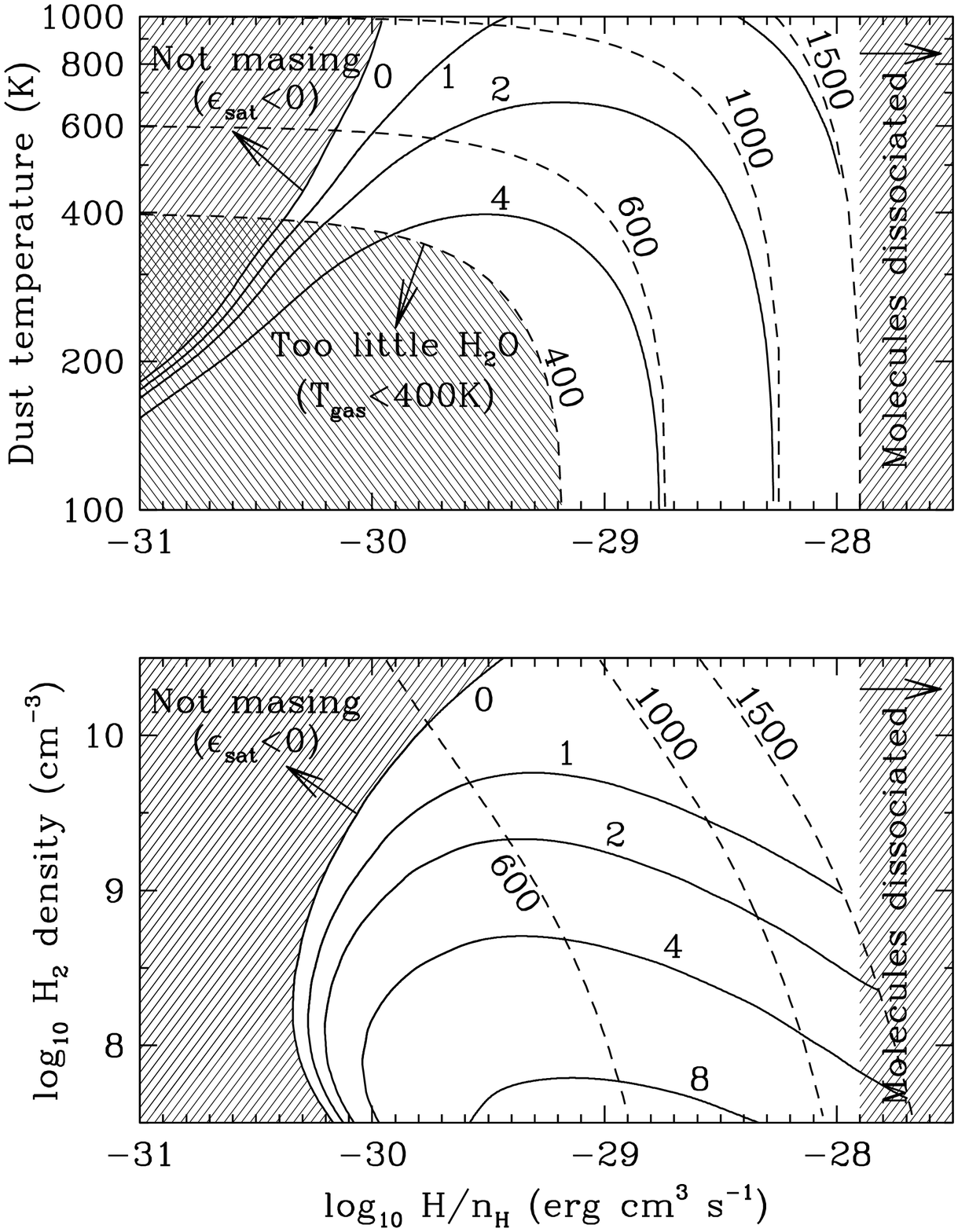}{300pt}{0}{50}{50}{-150}{-30}
\figcaption{Maser efficiency (solid contours labeled with
$10^5 \times \epsilon^0_{\rm sat}$) and gas temperature
(dashed contours labeled with $T_{\rm gas}$ in K).  Results
are shown as a function of heating rate, $H$, and either
dust temperature (Figure 1a, top panel, for which 
the H$_2$ density is fixed at $10^9\,\rm cm^{-3}$) 
or H$_2$ density (Figure 1b, bottom panel, for which
the dust temperature is fixed at 500~K).  The assumed
water abundance for both panels is $10^{-4}$ relative to H$_2$ .
The region favorable to
efficient maser emission is the unshaded region where 
the gas temperature is large enough ($\simgt 400$~K) to yield 
a substantial water abundance and where the heating rate supports
a value of ($T_{\rm gas} - T_{\rm dust}$) large enough to
permit maser action but is not so large as to dissociate
molecules.}
\end{figure}

\begin{figure}
\plotfiddle{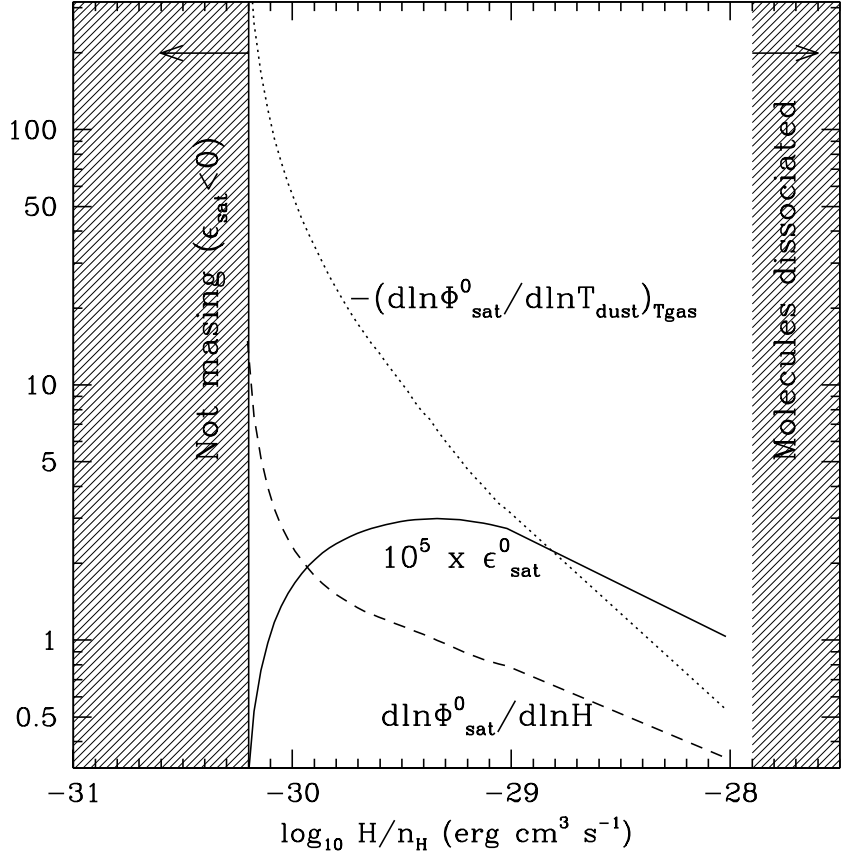}{300pt}{0}{60}{60}{-200}{-100}
\figcaption{Response of maser emissivity to small changes
in the heating rate and dust temperature, for 
a dust temperature of 500~K and a H$_2$ density of
$10^9\,\rm cm^{-3}$.  Values of the derivatives
$\partial \ln \Phi^0_{\rm sat} / \partial \ln H$
(dashed curve)
and $(\partial \ln \Phi^0_{\rm sat} 
/ \partial \ln T_{\rm dust})_{T_{\rm gas}}$ (dotted curve)
are shown as a function of the heating rate, $H$ (see text).
The solid curve shows the maser emissivity 
(value plotted is $10^5 \times \epsilon^0_{\rm sat}$.)}
\end{figure}

\begin{figure}
\plotfiddle{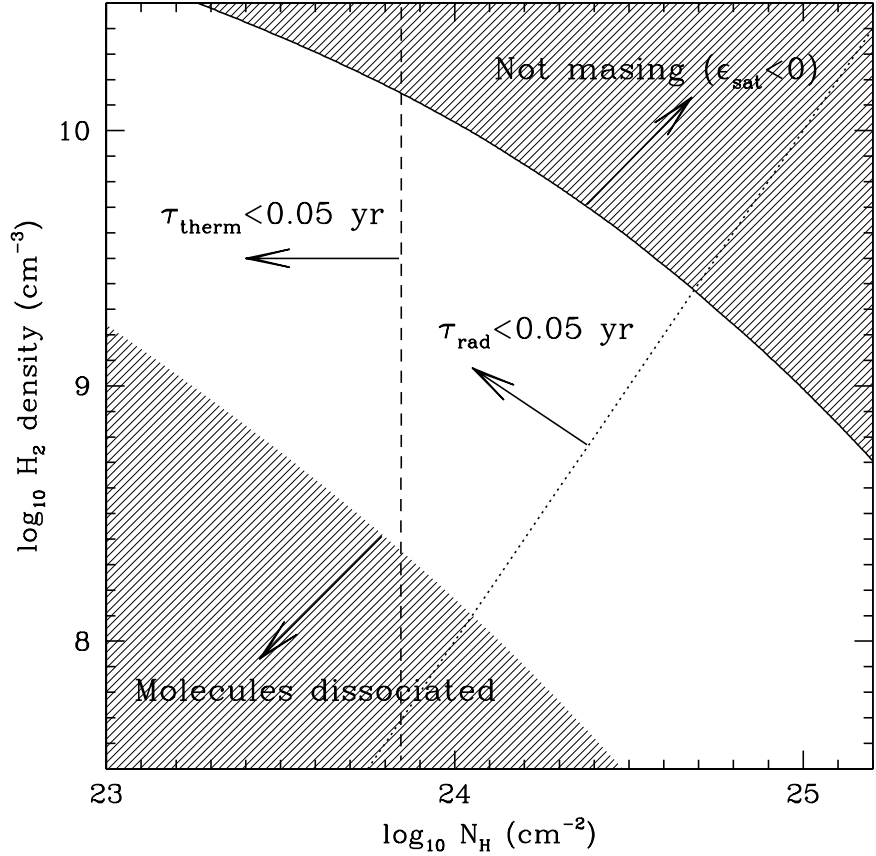}{300pt}{0}{60}{60}{-200}{-100}
\figcaption{Parameter space for maser action and
variability in clouds 0.7~pc distant from 
the active nucleus of NGC 1068.  Results are
shown as a function of the H$_2$ density
and the column density for X-ray shielding.
Regions of parameter space for which 
there is no maser emission are shaded.
The dashed line shows where the thermal timescale
for the gas is 0.05~yr, and the dotted line
shows where the radiative diffusion timescale
governing changes in the dust temperature is 0.05~yr.
The coordinated maser variability reported
by Gallimore et al.\ (2000) 
can be driven by changes in the X-ray
heating rate for parameters to the left of the
line $\tau_{\rm therm} = 0.05$~yr; or alternatively
by changes in the dust temperature for the region
to the left of the line $\tau_{\rm rad} = 0.05$~yr and to the
right of the line $\tau_{\rm therm} = 0.05$~yr.}
\end{figure}

\begin{figure}
\plotfiddle{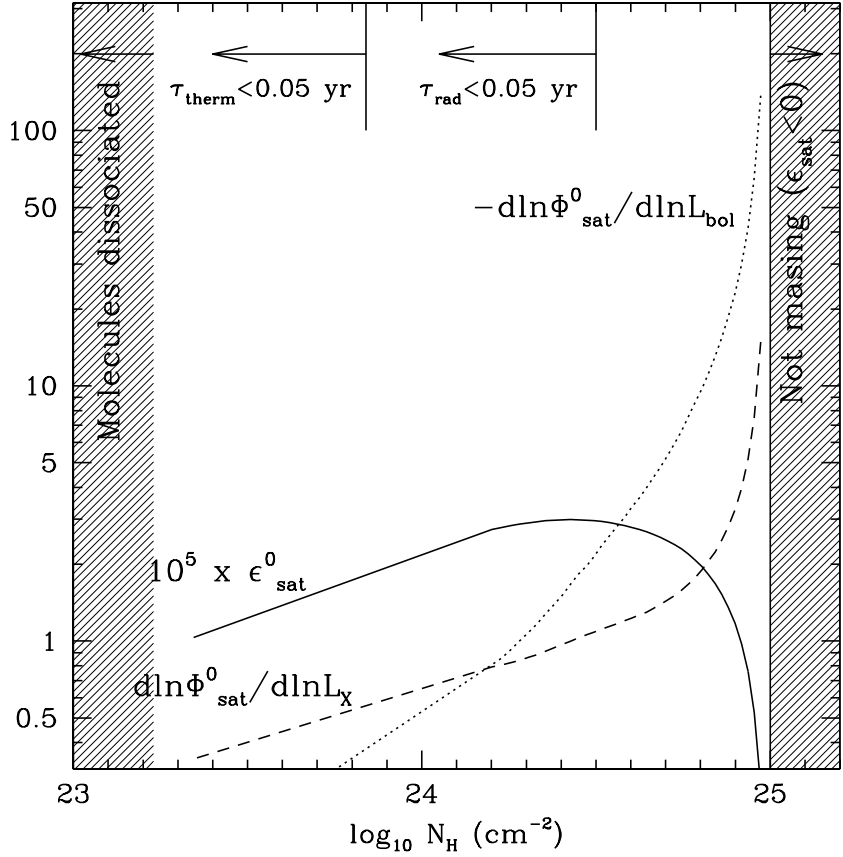}{300pt}{0}{60}{60}{-200}{-100}
\figcaption{Same as Figure 2, but now plotted as
a function of the column density for X-ray shielding.
Results apply to a cloud of H$_2$ density
$10^9\,\rm cm^{-3}$ that is 0.7~pc
distant from the active nucleus of NGC 1068.
Arrows indicate the range of column densities for
which the thermal timescale for the gas 
and the radiative diffusion timescale governing
changes in the dust temperature are smaller
than 0.05~yr.}
\end{figure}

\end{document}